\documentstyle[12pt]{article}
\newcommand{\be}{\begin{equation}}
\newcommand{\ee}{\end{equation}}
\newcommand{\bea}{\begin{eqnarray}}
\newcommand{\eea}{\end{eqnarray}}
\newcommand{\nn}{\nonumber}

\newcommand{\bra}[1]{\left\langle #1 \right|}
\newcommand{\ket}[1]{\left| #1 \right\rangle}

\begin{document}
\setlength{\baselineskip}{3.0ex}
\vspace*{1cm}
\rightline{BARI-TH/96-256}
\rightline{CPT-96/P.3412}
\rightline{SISSA 183/96/EP}
\rightline{UGVA-DPT/12/960}
\vspace*{2cm}
\begin{center}
{\large\bf Non leptonic heavy hadron decays and local duality\\}
\vspace*{6.0ex}
{\large\rm A. Deandrea$^a$, N. Di Bartolomeo$^b$, R. Gatto$^c$,
G. Nardulli$^d$\\}
\vspace*{1.5ex}
{$^a$\it Centre de Physique Th\'eorique, CNRS-Luminy, 13288 Marseille, France\\}
{$^b$\it Sissa, via Beirut 2-4, I-34014 Trieste, Italy\\}
{$^c$\it Dept. Physique Th\'eorique, Universit\'e de Gen\`eve, Switzerland\\}
{$^d$\it Dipartimento di Fisica, Univ. di Bari and INFN, Sez. Bari, Italy}
\end{center}
\noindent
\begin{quotation}
\vspace*{3cm}
\begin{center}
\begin{bf}
Abstract
\end{bf}
\end{center}
\vspace*{5mm}
\noindent
We discuss local duality for weak non leptonic $B$ and $\Lambda_b$ decays in
the $m_b\to \infty$ limit under the hypothesis of factorization and in the
Shifman-Voloshin limit. We show that, under these hypotheses, local duality
holds for the first two terms in the $1/m_b$ expansion and at the order
$\alpha_s$ in the perturbative expansion. The possible relevance of these
results for the operator product expansion evaluation of the $b$-hadron
lifetimes is discussed.
\end{quotation}
\newpage

\section{Introduction: inclusive heavy hadron decays and OPE}

In the last few years there has been an increasing
theoretical interest in inclusive heavy hadron
weak decays, mainly motivated by the possibility to use
the theoretical method of Operator Product Expansion (OPE) to
get information on the decay widths in the infinite heavy quark mass
limit: $m \to \infty$ \cite{bigi,falk}.
By employing OPE, a time ordered product of operators, in the present
case the weak hamiltonian responsible for the decay, is expanded in terms
of a sum of local operators of increasing dimension, multiplied by
inverse powers of the heavy-quark mass. In the $m \to \infty$
limit, one may truncate this expansion to the first few terms, that
are calculable in terms of perturbative QCD and some non perturbative
contributions that depend on some known matrix elements of
higher dimension operators.

This procedure can be illustrated by taking the semileptonic
inclusive decay width as an example. For a hadron $H_b$ containing
one heavy quark $b$ (e.g. $B$ or $\Lambda_b$) the inclusive width
is given by
\begin{equation}
\Gamma (H_b) = \frac{1}{2 m_H} disc {\bra {H_b}} {\cal{T}} {\ket {H_b}}
\label{1}
\end{equation}
where
\begin{equation}
{\cal{T}} = i \int d^4 x T({\cal{L}}(x) {\cal{L}}(0))
\label{2}
\end{equation}
and ${\cal{L}}$ is the effective lagrangian responsible for the transition. In
the following we consider the transition $b \to c$ only. For semileptonic
inclusive decays one has:
\begin{equation}
{\cal{L}}_{sl}=\frac{G_F}{\sqrt{2}} V_{cb} \bar{c} \gamma_{\mu} (1-\gamma_5 ) b
\bar{\ell} \gamma^{\mu} (1-\gamma_5 ) \nu_{\ell} = \frac{G_F}{\sqrt{2}} V_{cb}
J^{\mu}_h J_{\ell \mu}
\label{3}
\end{equation}
where $J^{\mu}_h$ and $J^{\mu}_{\ell}$ are the hadronic and leptonic currents
respectively.

As we mentioned already, the operator product expansion,
as applied to the product of operators
appearing in (\ref{2}), gives rise to an expansion in terms of decreasing powers
of the $b$-quark mass $m_b$;
for semileptonic decays the first terms of this expansion have
the form \cite{bigi}:
\begin{equation}
\Gamma_{sl}^{OPE} (H_b \to X_c) = \frac{G_F^2
m_b^5}{192 \pi^3} |V_{cb}|^2 \left[
R(x) \left( 1-\frac{K}{m_b^2} \right) + R'(x) \frac{G}{m_b^2} \right]
\label{10}
\end{equation}
where $x=m_c^2/m_b^2$ and
\begin{equation}
R(x)=1- 8x + 8x^3 -x^4 -12 x^2 \log x
\label{11}
\end{equation}
\begin{equation}
R'(x)=3 - 8x + 24 x^2 -24 x^3 +5 x^4 + 12 x^2 \log x.
\label{12}
\end{equation}
The leading term in (\ref{10})
corresponds to the leading $\bar{b} b$ operator in
the OPE, whereas the next-to-leading terms (of order ${\cal{O}} (1/m_b^2$))
arise
from the color magnetic moment operator and the $b$-quark kinetic energy; they
are proportional to the hadronic matrix elements
\begin{equation}
G = Z {\bra {H_b(v)}} \bar{b}_v \frac{g \sigma^{\mu \nu} G_{\mu \nu}}{4} b_v
{\ket {H_b(v)}}
\label{13}
\end{equation}
\begin{equation}
K = - {\bra {H_b(v)}} \bar{b}_v \frac{(i D)^2}{2} b_v {\ket{H_b(v)}} .
\label{14}
\end{equation}
Here $b_v$ is the effective, velocity dependent heavy quark operator of the
heavy quark effective theory (HQET), $Z$ is a renormalization factor equal to
one at the scale $\mu =m_b$. In the $m_b \to \infty$ limit, $G$ and $K$ are
finite. Let us incidentally note that,
for the leading term in (\ref{10}),
corresponding to the parton model result
(i.e. (\ref{10}) without the corrections of the order ${\cal{O}} (1/m_b^2$)),
also the
${\cal{O}}(\alpha_s)$ corrections are available \cite{falk}, i.e.
\begin{equation}
\Gamma_{sl} (b \to c \ell \nu_{\ell}) =
\frac{G_F^2 m_b^5}{192 \pi^3} |V_{cb}|^2
\left[ R(m_c^2/m_b^2) + \frac{\alpha_s}{\pi} A_0(m_c^2/m_b^2) \right]
\label{15}
\end{equation}
where $A_0(x)$ contains the perturbative corrections and can be found in
\cite{falk}.

The idea that OPE can be applied to heavy hadron decays is based on the
notion of quark-hadron duality, a concept which is widely used, but remains
nonetheless rather vague. The
physical meaning of duality is intuitively clear: because of the large
$b-$quark mass, the energy release to the final hadrons is large as
compared to $\Lambda_{QCD}$; therefore one expects that QCD can be
applied and this would justify the approximation of identifying
the hadronic decay width with the corresponding OPE expression.

However a closer scrutiny shows that this expectation is
in general false and can be retained only if some further assumptions are
made. The reason why the physical and the OPE widths cannot
be identical is in the different structure of their
singularities: in one case there are multiparticle
hadronic final states and, therefore, multihadron thresholds, while
in the other case one has quark and gluon production with a different
set of thresholds.

There are two possibilities for avoiding this negative conclusion. The first
one consists in working far away from the physical thresholds; in this case
OPE becomes necessarily true and one can therefore safely compute the
hadronic quantities by their simpler OPE counterparts. However this
local parton-hadron duality does not hold in semileptonic $H_b\to X_c
\ell \nu_\ell$ weak decays (neither it occurs
in other high energy processes, e.g. $e^+ e^- \to$ hadrons). The reason for
this can be easily understood by looking at the differential decay rate
${d \Gamma}/{dq^2dydv{\hat q}}$
(here $q=p_\ell +p_\nu,~y={E_\ell}/{m_b},~p_b^\mu=m_bv^\mu,~
v{\hat q}={vq}/{m_b}$). For $q^2$ and $y$ fixed, the
differential decay rate, in the complex $v{\hat q}$ plane, has a cut,
corresponding to multihadron production. For OPE to be valid, one should be
away from the cut, which is not the case. The second possibility,
that can be applied in semileptonic
$H_b$ decays, consists in considering, instead of the hadronic and OPE
expressions, some averages, i.e. their smearing over a region of
the order $\Lambda_{QCD}$ \cite{wein}. The use of smeared quantities is
generally referred to as global duality and has been adopted in \cite{boyd}
to prove the validity of OPE for semileptonic $B$ and $\Lambda_b$ decays
in a particular kinematical region, the Shifman-Voloshin (SV) limit \cite{sv}
(see below).
Considering the integrated semileptonic width, the authors in
\cite{boyd} show that OPE is valid as well, even without
explicit smearing. The matching between
the hadronic and the OPE expressions for
the widths is proved to two orders in the $1/m$ expansion
and to first order in $\alpha_s$. As already suggested in \cite{dun},
the validity of OPE is
due to the fact that the integration contour can be deformed
in such a way that the part which remains near the cut is of order
$\Lambda_{QCD}/m$ and therefore is vanishingly small
in the $m \to \infty$ limit.

The validity of OPE for non-leptonic decays is less clear.
For non-leptonic transitions, neglecting penguin
operators and four quarks $b \bar{c} \bar{c} s$ operators, we have the
effective lagrangian:
\begin{equation}
{\cal{L}}_{nl}=\frac{G_F}{\sqrt{2}} V_{cb} V^*_{ud} [ c_1 O_1 + c_2 O_2 ].
\label{4}
\end{equation}
Here $c_1$ and $c_2$ are Wilson coefficients that in the leading-log
approximation are given by:
\bea
c_1&=&\frac{1}{2}
\left(\left[ \frac{\alpha_s (m_W)}{\alpha_s (m_b)} \right]^{6/23}  +
\left[ \frac{\alpha_s (m_W)}{\alpha_s (m_b)} \right]^{-12/23}\right)
\nn \\
c_2&=&\frac{1}{2}
\left(\left[ \frac{\alpha_s (m_W)}{\alpha_s (m_b)} \right]^{6/23}  -
\left[ \frac{\alpha_s (m_W)}{\alpha_s (m_b)} \right]^{-12/23}\right)
\label{5}
\eea
and the operators $O_j$ are given by:
\begin{eqnarray}
O_1 &=& \bar{c} \gamma_{\mu} (1-\gamma_5) b \bar{d} \gamma^{\mu} (1-\gamma_5)
u \label{8} \\
O_2 &=& \bar{c} \gamma_{\mu} (1-\gamma_5) u \bar{d} \gamma^{\mu} (1-\gamma_5)b
~~.\label{9}
\end{eqnarray}

The OPE expression for
the non leptonic decay width can be written in terms of the
semileptonic  $\Gamma_{sl}^{OPE}$ as follows \cite{stech}:
\bea
\Gamma_{nl}^{OPE} & = & N_c  |V_{cb}|^2
\Gamma_{sl}^{OPE} \left\{ (c_1^2 + c_2^2)
\left(1 + \frac{\alpha_s}{\pi} \right) + \right. \nn \\
& + & \left.  \frac{2 c_1 c_2}{N_c} \left[1 - \frac{16}{3} \frac{\alpha_s}{\pi}
+\frac{16}{m_b^2} \frac{(1 - m_c^2/m_b^2)^3}{R(m_c^2/m_b^2)} G \right]
\right\}~.
\label{16}
\eea
In this case, however, we have {\it a priori} no reason to expect that the
contribution near the physical cut is small, because there is no external
momentum $q$ and no integration contour to deform. Therefore,
for non leptonic decays, in order to justify OPE, one has to {\it assume}
local duality or, alternatively,
to prove it under some additional hypothesis. Clearly this puts the
validity of OPE for non leptonic heavy hadron decays on a less
firm ground \cite{dun}; a signal of this might be the experimental
result for the ratio of the total lifetimes of $\Lambda_b$ and $B$.
Experimentally one has ${\tau(\Lambda_b)}/{\tau(B)}=
0.76 \pm 0.06$ (see e.g. \cite{exp}), whereas, from OPE, one expects
${\tau(\Lambda_b)}/{\tau(B)}\simeq 1$ (including corrections to the third
order in the $1/m$ expansion \cite{fdfz}). As an explanation of this
discrepancy, a possible failure of local duality for non leptonic heavy
hadron decays has been suggested \cite{alt}.

The aim of this paper is to investigate the validity of OPE
for non leptonic $B$ and $\Lambda_b$ decays. We shall prove that
local duality is indeed valid and that OPE can be proved
under the following hypotheses:

1) one works in the SV  \cite{sv} limit:
\be
m_b, ~m_c ~ >> \delta m = m_b - m_c ~ >> \Lambda_{QCD}
\label{b1}
\ee

2) the weak non-leptonic amplitude is factorized in
two parts, the first one corresponding to the transition $b\to c$
and the second to a transition between light quarks. Since 
factorization has been proved only in the $N_c \to \infty$ limit
($N_c=$ number of colours), we shall implicitly assume this limit,
even if we retain some of the ${\cal O}(1/N_c)$ corrections
in order to keep track of terms violating factorization and/or duality.

We shall show in the next section how these hypotheses are
used to prove duality while in Section 3 we shall conclude this letter
by a discussion of our results.

\section{$B$ and $\Lambda_b$ non leptonic decays and duality}

Before considering the problem of local duality for non-leptonic $B$ and
$\Lambda_b$ decays, we shall briefly review the results obtained in
\cite{boyd} for the semileptonic case.

These authors work in SV limit (\ref{b1}).
The OPE result for semileptonic decays, as expressed by (\ref{10})
(with the corrections ${\cal O} (\alpha_s )$ in (\ref{15})), becomes,
in this  limit:
\bea
\Gamma_{sl}^{OPE} (H_b \to X_c) &=& \frac{G_F^2}{192 \pi^3} |V_{cb}|^2
\left[ \left( 1 - \frac{\alpha_s}{\pi} \right) \left( \frac{64}{5} \delta m^5
 - \frac{96}{5} \frac{\delta m^6}{m_b} \right) \right. \nn \\
&+& \left. 64 \frac{G \delta m^4}{m_b} + \ldots \right]
\label{b2}
\eea
The dots indicate several higher order corrections, i.e. terms ${\cal O}
(\delta m^7/m_b^2 )$ or  ${\cal O} ( G \delta m^5/m_b^2 )$ or
${\cal O} ( K \delta m^5/m_b^2 )$; they can be found in \cite{boyd}.
As observed in \cite{boyd} a correction ${\cal O} ( G \alpha_s/m_b )$
could also be added but has not been computed yet (note that $G=0$ when
$H_b = \Lambda_b$ because the light degrees of freedom have zero angular
momentum). Computing now $\Gamma_{sl} ( B \to D,D^* l \nu)$ and $\Gamma_{sl}
( \Lambda_b \to \Lambda_c l \nu)$ in the same kinematical regime (\ref{b1}),
\footnote{Under the SV kinematical conditions, the form factors can be
expanded around the zero recoil point and corrections to this limit appear
only at the order $1/M^2$.} one obtains the results \cite{boyd}:
\bea
\Gamma_{sl}^{had} (B \to D,D^* \ell \nu_{\ell} ) & =  &
\frac{G_F^2}{192 \pi^3} |V_{cb}|^2
\left[ \left( 1 - \frac{\alpha_s}{\pi} \right) \left( \frac{64}{5} \delta M^5
 - \frac{96}{5} \frac{\delta M^6}{M_B} \right) \right. \nn \\
& +& 64
\left. \left( 1 - \frac{4 \alpha_s}{3 \pi} \right) \frac{G \delta M^4}{M_B}
 + \ldots \right]
\label{b3}
\eea
with
\be
\delta M = M_B - M_D = \delta m \left[ 1 - \frac{K_B + G}{m_b^2}\right] +
{\cal{O}}(1/m_b^3)
\label{b4}
\ee
and
\be
\Gamma_{sl}^{had} (\Lambda_b \to \Lambda_c
\ell \nu_{\ell})  =  \frac{G_F^2}{192 \pi^3} |V_{cb}|^2
\left[ \left( 1 - \frac{\alpha_s}{\pi} \right) \left( \frac{64}{5} \delta M^5
 - \frac{96}{5} \frac{\delta M^6}{M_{\Lambda_b}} \right)
 + \ldots \right]
 \label{b5}
 \ee
with
\be
\delta M = M_{\Lambda_b}
- M_{\Lambda_c} = \delta m \left[ 1 - \frac{K'}{m_b^2}\right] +
{\cal{O}}(1/m_b^3)~~.
\label{b6}
\ee
$K_B$ and $K'$ in (\ref{b4}) and (\ref{b6}) are given by (\ref{14}) with
$H_b = B$ and $H_b = \Lambda_b$ respectively.
The inclusive semileptonic width for $B$ decay
is saturated, in the SV limit, by decays
into $D$  and $D^*$ and, for $\Lambda_b$, by its semileptonic decay into
$\Lambda_c$.

Eqs. (\ref{b2}), (\ref{b3}) and (\ref{b5}) show that, for semileptonic
inclusive decays, duality holds in the SV limit for the first two terms in
$1/m_b$ expansion and at the first order in $\alpha_s$; the only term
for which duality has not been checked in \cite{boyd} is
the term proportional to $\alpha_s G \delta M^4/M_b$ appearing in
(\ref{b3}), which has no matching in (\ref{b2}) (as we stressed
already, the calculation of
this correction to the OPE leading term is still missing).

Let us show how the OPE inclusive non-leptonic decay rate (\ref{16}),
can be matched by the sum over exclusive  non-leptonic
decay channels. Similarly to the case of semileptonic decays,
we shall work in the SV limit and, therefore, we shall take the results
of \cite{boyd}, i.e. that duality is proved for the first two terms in the
$1/m_b$ expansion and at the first order in $\alpha_s$ for the semileptonic
decays. As a second hypothesis, we shall assume factorization of the decay
amplitudes. Because of our hypotheses, we expect that, also for non leptonic
decays, duality may be proved only for the first two leading terms in
$1/m_b$ expansion and up to order $\alpha_s$.

The full width for non-leptonic decays is given by
\bea
\Gamma_{nl} & = & \frac{1}{2 M_{H_b}} ~disc \left\{
i \int d^4 x {\bra{H_b (v)}} T( {\cal L}_{nl} (x) {\cal L}_{nl} (0) )
{\ket{H_b (v)}} \right\} \nn \\
& = & \frac{1}{2 M_{H_b}} \sum_{X} (2 \pi)^4 \delta^4 (p - p_X ) \left|
{\bra{X}} {\cal L}_{nl} (0) {\ket{H_b (v)}} \right| ^2
\label{n2}
\eea
where ${\cal L}_{nl}$ is the effective lagrangian for non leptonic decays
given in (\ref{4}).

Let us now write
\be
{\ket{X}} = {\ket{X_u X_c}}
\label{n3}
\ee
where $X_u$ is a set of light particles, and $X_c$ is a set of hadronic
charmed states. Let us now consider the matrix element
\be
{\bra{X_u X_c}} {\cal L}_{nl} (0) {\ket{H_b (v)}}~~.
\label{n4}
\ee

Assuming factorization, i.e. inserting the vacuum in all possible ways,
we obtain, by using the Fierz identities,
\bea
{\bra{X_u X_c}} {\cal L}_{nl} (0) {\ket{H_b (v)}} & = &
\left( c_1 + \frac{c_2}{N_c} \right)
\frac{G_F}{\sqrt{2}} V_{cb} V_{ud}^{*}{\bra{X_u}} {\bar d} \gamma^\mu
(1- \gamma_5) u {\ket{0}} \times \nn \\
& & {\bra{X_c}} {\bar c} \gamma^\mu (1- \gamma_5) b{\ket{H_b (v)}}.
\label{n5}
\eea

A comment on the factorization procedure we have adopted is now in
order.
~From eqs.(\ref{4}),(\ref{8}) and (\ref{9}), we see that,
using the Fierz theorem, one can write for ${\cal{L}}_{nl}(0)$
the following expressions:
\begin{equation}
{\cal{L}}_{nl}(0)
=\frac{G_F}{\sqrt{2}} V_{cb} V^*_{ud} [ (c_1  + \frac{c_2}{N_c})
O_1~+~{\tilde O}_1 ]
\label{ac4}
\end{equation}
or
\begin{equation}
{\cal{L}}_{nl}(0)=
\frac{G_F}{\sqrt{2}} V_{cb} V^*_{ud} [ (c_2  + \frac{c_1}{N_c})
O_2~+~{\tilde O}_2 ] ~~.
\label{ac5}
\end{equation}
${\tilde O}_1$ and $ {\tilde O}_2 $ are written as products of two
coloured currents; in the factorization approximation
they give no contribution because the matrix elements of a coloured current
between physical hadronic states vanish. In exclusive decays,
(\ref{ac4}) and  (\ref{ac5}) produce respectively the so-called {\it class I}
and {\it class II} decays. We shall now show that, within our hypotheses,
we can always choose the form (\ref{ac4}) for ${\cal{L}}_{nl}$. In order
to be definite let us consider $H_b={\bar B}^0$ decays. If
$\ket {X_u X_c}$ contains $D^+$ or $D^{*+}$ states, we can assume
eq.(\ref{ac4}) and apply factorization as in (\ref{n5}), with
$\ket{X_c}=\ket{D^+}$ or $\ket{D^{*+}}$ and neglecting ${\tilde O}_1$.
If $\ket{X_u X_c}$ contains $D^0$ or
$D^{*0}$ states, we can evaluate the
matrix element in two ways.

1) We can write
${\cal{L}}_{nl}(0)$ according to (\ref{ac4}) and neglect
${\tilde O}_1$. In this case one shall factorize the amplitude as
follows (considering only $D^0$ for simplicity):
\bea
{\bra{X_u D^0}} {\cal L}_{nl} (0) {\ket{{\bar B}^0}} & = &
\left( c_1 + \frac{c_2}{N_c} \right)
\frac{G}{\sqrt{2}} V_{cb} V_{ud}^{*}{\bra{X^-}} {\bar d} \gamma^\mu
(1- \gamma_5) u {\ket{0}} \times \nn \\
& & {\bra{X^+D^0}} {\bar c} \gamma^\mu (1- \gamma_5) b{\ket{H_b (v)}}.
\label{ac6}
\eea
where $X^+,X^-$ are multiparticle states containing only light hadrons
(the superscript $\pm$ indicates the total charge). On the basis of the
results for semileptonic decays contained in \cite{boyd} and \cite{sv},
(\ref{ac6}) represents in the SV limit a higher order
 correction (of the order of
${\cal O}({\delta M^2}/{M_b^2})$) because ${\bar B}^0$ decays
semileptonically only to $D^+,D^{*+}$ in this limit.

2) The second way to evaluate
the contribution of these states is by making use of (\ref{ac5})
and neglecting, in the factorization approximation, ${\tilde O}_2$. However
one can see immediately that, also by this choice, the matrix element
is a higher order correction and should be neglected. As a matter of fact,
since we work in the $N_c \to \infty$ limit, where factorization
can be proved to be valid, we should evaluate the resulting expression
in this limit (for discussion on this point see \cite{s4}, \cite{alek}). But
in the $N_c \to \infty$ limit these states would contribute to the width
a term ${\cal O}(c_2+c_1/N_c)^2= {\cal O}(c_2)^2=
{\cal O}(\alpha_s^2)$; in the approximation when one neglects
higher order ${\cal O}(\alpha_s^2)$ perturbative contributions, this term should
be neglected as well.

On the basis of this discussion we conclude that, in the SV limit,
and taking into account only the first two terms in
the $1/m_b$ expansion  and at order $\alpha_s$ in the perturbative series,
we can limit ourselves to consider only {\it class I} decays,
as expressed by (\ref{n5}).

~From (\ref{n2}) and (\ref{n5}) one obtains
\be
\Gamma_{nl} (H_b) = \frac{G_F^2}{2} |V_{cb} V^*_{ud}|^2
\left( c_1 + \frac{c_2}{N_c} \right)^2 \int \frac{d^4 q}{(2 \pi)^3}
W_{H_b}^{\mu \nu} (q,v) T_{\mu\nu}(q),
\label{n6}
\ee
where $W_{H_b}^{\mu\nu}$ is  the same hadronic tensor appearing in
the semileptonic  $b \to c$ decays
\bea
W_{H_b}^{\mu\nu} & = & (2 \pi )^3 \sum_{X_c} \delta^4 ( p_{H_b} - q - p_{X_c})
\times \nn \\
& & {\bra{H_b}} {\bar b} \gamma^\mu (1- \gamma_5) c{\ket{X_c}}
{\bra{X_c}} {\bar c} \gamma^\nu (1- \gamma_5) b{\ket{H_b}}
\label{n7}
\eea
while $T_{\mu \nu}$ is given by:
\bea
T_{\mu\nu} (q) & = & (2 \pi )^4 \sum_{X_u} \delta^4 (q - q_{X_u})
\times \nn \\
& &
{\bra{0}} {\bar u} \gamma_\mu (1- \gamma_5) d{\ket{X_u}}
{\bra{X_u}} {\bar d} \gamma_\nu (1- \gamma_5) u{\ket{0}}~~.
\label{n8}
\eea

The vector part of the tensor $T_{\mu \nu}$ is theoretically
known for large values of $q^2$ up to  the order $\alpha_s^3$ from
$e^+ e^-$ into hadrons: here we need only the expression to the first
order in $\alpha_s$. We shall write
\be
T^{\mu\nu} ( q) = T^{\mu\nu}_R ( q) + T^{\mu\nu}_{QCD} ( q)~~,
\ee
where $ T^{\mu\nu}_{QCD} (q)$ is the $QCD$ contribution
to $T^{\mu\nu} ( q) $; neglecting the $u$ and $d$ quark masses,
it is given by:
\be
T^{\mu\nu}_{QCD} (q) = 2 N_c \frac{1}{6 \pi} \left( 1 + \frac{\alpha_s}{\pi}
\right)
( q^\mu q^\nu - q^2 g^{\mu \nu} ) \theta( q^2 - q_0^2)~~.
\label{n9}
\ee
The factor $2$ with respect to the result for $R_{e^+ e^-}$ is due to the
contribution of both vector and axial currents in (\ref{n8});
$q_0^2 \sim 1-2~GeV^2$ is the onset of the $QCD$
behaviour. $T^{\mu\nu}_R ( q)$ takes into account
the contribution of the low lying
($m^2_R \leq q_0^2$) particles: $\pi$, $a_1$, $\rho$, $\rho'$.
It can be easily checked that,
in the SV limit, these resonances contribute to the non-leptonic rate with
terms of the order $G_F^2 \delta m^3 f_R^2$, where $f_R$ is
a mass parameter, of order $\Lambda_{QCD}$, characteristic
of the resonance. As we shall discuss below, these terms represent corrections
${\cal O}({\Lambda_{QCD}}/{\delta m})^2$ to
the leading contribution arising from
$T^{\mu\nu}_{QCD} $ and will be neglected  because,
in the SV limit, $\delta m >> \Lambda_{QCD}$ (see eq. (\ref{b1})).
We shall therefore put
\be
T^{\mu\nu} (q) \simeq T^{\mu\nu}_{QCD}(q)   ~~.
\ee
To obtain the non leptonic decay width $\Gamma_{nl} (H_b)$
from (\ref{n6})
we observe that for the semileptonic inclusive $b \to c$ decay, the
following formula holds
\be
\Gamma_{sl} (H_b) = \frac{G_F^2}{2} |V_{cb}|^2
\int \frac{d^4 q}{(2 \pi)^3}
W_{H_b}^{\mu \nu} (q,v) T_{\mu\nu}^{l}(q)~~,
\label{n10}
\ee
where, for
$q^2>q_0^2$ ,
the leptonic tensor $T_{\mu \nu}^{l}$ differs
\footnote{
The presence of $ q^2_0\neq 0$ in (\ref{n9}) introduces corrections
${\cal O}({\Lambda_{QCD}^2}/{\delta m^2})$, as it can be easily
verified. Similarly to our discussion
above, we neglect this correction in the SV limit. } from (\ref{n9}) only by the
overall factor $N_c (1 + \alpha_s/\pi)$:
\be
T_{\mu \nu}^{l} (q) = 2  \frac{1}{6 \pi}
( q_\mu q_\nu - q^2 g_{\mu \nu} )~~.
\label{n11}
\ee

~From the previous formulae one obtains an expression for the non leptonic
width,
computed in the factorization approximation, as a function of
the semileptonic one:
\be
\Gamma_{nl}^{had}(H_b) = N_c \left(1 + \frac{\alpha_s}{\pi}
\right) |V_{cb}|^2
\left(c_1 + \frac{c_2}{N_c} \right)^2 \Gamma_{sl}^{had}(H_b) + ...
\label{n12}
\ee
where the omitted terms
are ${\cal O} \left({\Lambda^2}/{\delta m^2}\right)$ or ${\cal O} \left(
{\delta m^2}/{m_b^2}\right)$.
In the SV limit we can substitute in this equation expressions (\ref{b3})
and (\ref{b5}) for $B$ and $\Lambda_b$ respectively.

To compare this result with (\ref{16}), we develop the Wilson coefficients
$c_1$ and $c_2$ to the order $\alpha_s$:
\bea
c_1 (\mu) & = & 1 - \frac{3}{N_c} \frac{\alpha_s (M_W)}{4 \pi}
 \log(\mu^2/M_W^2) \nn \\
c_2 (\mu) & = & 3 \frac{\alpha_s (M_W)}{4 \pi}
 \log(\mu^2/M_W^2) ~~.
 \label{n13}
\eea
We notice that the combination $c_1 + c_2/N_c$ in (\ref{n12}), does not contain
first order $\alpha_s$ corrections.
By developing (\ref{16}) in the SV limit, we obtain, at the first order in
$\alpha_s$:
\bea
\Gamma_{nl}^{OPE} &  = & \frac{G_F^2 N_c}{192 \pi^3} |V_{cb}|^2
\left[ \frac{64}{5} \delta m^5
 - \frac{96}{5} \frac{\delta m^6}{m_b} + 64 \frac{G \delta m^4}{m_b}
 + \ldots \right] \nn \\
 & \times & \left( 1 + \frac{ 20 c_1 c_2}{N_c} \frac{ G}{\delta m^2} \right)
 \label{gope}~~.
 \eea

On the other hand, from (\ref{n12}), after substitution of (\ref{b3}) or
(\ref{b5}) we have the expression
\be
\Gamma_{nl}^{had} (H_b)  =  \frac{G_F^2 N_c}{192 \pi^3} |V_{cb}|^2
\left[ \frac{64}{5} \delta M^5
 - \frac{96}{5} \frac{\delta M^6}{M_{H_b}}  + 64 \frac{G \delta M^4}{M_{H_b}}
 + \ldots \right]
  \label{gope1}
\ee
which is valid in the same limit. Comparing these two equations, we see
that, a part from the term ${ 20 c_1 c_2 G}/({N_c \delta m^2})$, which
is present in (\ref{gope}) and has no counterpart in
(\ref{gope1}), there is matching between the two expressions.
We notice that, in this term, $G$ is of the order $\Lambda_{QCD}^2$;
therefore it represents a correction of the order $\alpha_s
\Lambda_{QCD}^2 \delta m^3$ and therefore belongs to
a class of corrections that we have neglected (for example those
coming from the low-lying resonances)\footnote{In this approximation
one should neglect the term $64 G \delta m^4/m_b$ as well.
It can be noted, however, that this term has an exact counterpart in
the hadronic expression.}. Moreover it represents a correction
${\cal O}(1/N_c)$ and should be neglected, in the approximation we
are considering.

The previous analysis shows
that the inclusive non leptonic  $B$ and $\Lambda_b$ decay widths,
computed in the Shifman-Voloshin limit (\ref{b1})  by factorization,
are equal to their OPE expressions
including the leading  term ${\cal O}(\delta m^5)$
and the next-to-leading term ${\cal O}({\delta m^6}/{m_b})$
in the OPE, at the order $\alpha_s$ in the perturbative expansion and
up to terms  ${\cal O}({\Lambda_{QCD}^2}/{\delta m^2})$
that are negligible in the SV limit. Within this limitations
our analysis shows that local duality is indeed satisfied by
non leptonic heavy hadron decays.

\section{Conclusions}

Let us summarize and discuss our results. We have proved local duality for
$B$ and $\Lambda_b$ non leptonic decays, i.e. the equalities
\be
\Gamma_{nl}^{OPE} (H_b)=  \Gamma_{nl}^{had}(\Lambda_b)=\Gamma_{nl}^{had}(B)~,
\label{n16}
\ee
for the first two terms in the $1/m_b$ expansion and at the order $\alpha_s$
in the perturbative expansion. These results
have been proved in the Shifman-Voloshin limit and
assuming factorization of the weak amplitudes.

One may wonder on the experimental significance of SV limit; for example, in
this kinematical regime, the $B$ meson would decay semileptonically
only to $D$ and $D^*$. This prediction
however, does not agree with the data, since experimentally one has
\be
{\frac{\Gamma (B \to D,D^* \ell \nu_{\ell} )}{
\Gamma  (B \to X \ell \nu_{\ell} )} = 0.64 \pm 0.08}
\ee
which is obtained considering both $B^0$ and $B^\pm$ semileptonic
decays \cite{pdb}. This discrepancy is expected since
the assumption (\ref{b1}) $m_c >> \delta m$ is clearly violated.

Secondly, our results are based on the assumption of
factorization of the hadronic final state into two parts, one
containing a single charmed particle ($D, D^*$ or $\Lambda_c$) and
the other containing only light hadrons. Factorization
holds in the $N_c \to \infty$ limit, and can therefore be violated
by terms ${\cal O}(1/N_c)$. Moreover, there are arguments that
can justify factorization in the $m_b \to \infty$ limit \cite{dug},
but it is not clear to which order in the $1/m_b$ expansion it
holds. In \cite{blok}, an estimate is given of the nonfactorizable part
of the exclusive process $B^0 \to D^+ \pi^-$: they found a term
proportional to the matrix element $G$ (see eq. (\ref{13}))
and color-suppressed with respect to the leading one;
moreover these authors find that in the SV limit such a term
is not suppressed by powers of $1/m_b$. This result agrees
with our eq. (\ref{gope}), where similar terms are present
(those proportional to ${ 20 c_1 c_2 G}/({N_c \delta m^2})$).
As discussed above, besides being colour suppressed,
in the SV limit these contributions are negligible because are of
order ${\Lambda^2_{QCD}}/{\delta m^2}$; it is quite
possible that, beyond the SV limit, nonfactorizable terms arise that
are not suppressed by $1/m^2_b$ in the infinite heavy quark mass limit.
In \cite{blok2} the problem of factorization in inclusive non
leptonic $B$ decays is discussed assuming a standpoint complementary
to the one taken in the present paper: these authors
assume local duality from the outset and discuss the limits of factorization.
As far as our approach is comparable with theirs, the results
of the present paper agree with those of \cite{blok2}.

Finally let us note that,
because of the assumed SV limit, the non leptonic
widths depend on $\delta M$ and not on $M_{H_b}$. Because
of (\ref{b4}) and (\ref{b6}), both for $B$ and $\Lambda_b$
decays $\delta M = \delta m+ {\cal O}(1/m_b^2)$ and there
is no way for a ${\cal O}(1/m_b)$ term to appear in the ratio
$\tau(\Lambda_b)/\tau(B)$, contrarily to the possibility
suggested in \cite{alt}. This result, however, might be
heavily dependent on the kinematics of the SV limit and
the possibility remains that, beyond this limit,  ${\cal O}(1/m_b)$
corrections to the non leptonic $B$ and $\Lambda_b$ lifetimes appear.

\par\noindent
\vspace*{1cm}
\par\noindent
{\bf Acknowledgements}
\par\noindent
This work was carried out under the program Human Capital and Mobility, 
contract nr. ERBCHRXCT940579, OFES nr.950200.
A.D. acknowledges the support of a TMR research fellowship of the European
Commission under contract nr. ERB4001GT955869.  We thank F. Feruglio for
reading of the manuscript.

\end{document}